\documentclass[aps,pra,floatfix,showpacs,preprintnumbers,amsmath,amsymb,a4paper,superscriptaddress,twocolumn,longbibliography]{revtex4-1}
\usepackage{amsmath}
\usepackage{graphicx}
\usepackage{amssymb}
\usepackage{amsmath}
\usepackage{epstopdf}
\usepackage{float}
\usepackage{color}

\newcommand{\be}{\begin{equation}}
\newcommand{\ee}{\end{equation}}

\begin{document}

\title{Formation of long-range Rydberg molecules in two-component ultracold gases}

\author{Matthew T. Eiles}

\affiliation{Department of Physics and Astronomy, 
Purdue University, 47907 West Lafayette, IN, USA}
\affiliation{Max-Planck-Institut f\"{u}r Physik komplexer Systeme, N\"{o}thnitzer Str. 38, 01187 Dresden, Germany}

\date{\today}

\begin{abstract}
We present a comprehensive study of the diverse properties of heteronuclear Rydberg molecules, placing a special emphasis on those composed of the light alkali atoms, Li, Na, and K. Electron-atom scattering phase shifts, which determine the strength of the molecular bond, are calculated at very low energy and then used in a spin-dependent theoretical model to calculate accurate Rydberg molecule potential energy curves. The wide parameter range accessible by combining the various properties of different alkali atoms often leads to hybridized electronic states accessible via one or two photon excitation schemes. This analysis of heteronuclear molecules leads to a prediction that the relative densities and spatial distributions of atoms in an ultracold mixture can be probed at controllable length scales via spectroscopic detection of these molecules.
\end{abstract}

%\keywords{}

\maketitle
\section{Introduction}
Long-range Rydberg molecules have attracted much interest since their experimental observation nearly a decade ago \cite{Bendkowsky}. Subsequent experiments observed all the molecular states originally predicted \cite{Greene2000,HamiltonGreeneSadeghpour,KhuskivadzeJPB}, from the low-$l$ non-polar Rydberg molecules \cite{quantumreflection,AndersonPRL,Sass} to the exotic polar ``trilobite'' and ``butterfly'' states \cite{Butterfly,BoothTrilobite}. These observations were sophisticated enough to require careful studies of fine and hyperfine structure and singlet/triplet scattering symmetries. These molecules have been observed in three atomic species (Rb, Cs, and Sr) and proposed in several others (Mg, Ca, and Li) \cite{Butterfly,AndersonPRL,EilesSpin,BoothTrilobite,quantumreflection,Sass,DeSalvo2015,BoothTrilobite,Eiles2015,NewPfau,BartschatSadeghpour}.  
Tantalizing field control opportunities, stemming from the huge dipole moments of these molecules, have inspired theoretical \cite{EilesPendular,KurzSchmelcherPRA,Hummel} and experimental \cite{PfauKurz,GajKrupp,Butterfly} studies. Finally, the observation of polyatomic Rydberg molecules in dense atomic clouds and at high principal quantum numbers has sparked interest, and remains a challenging theoretical problem in all but the simplest Rydberg states \cite{Rost2006,JPBdens,EilesHyd,FeyKurz,Demler,WhalenPoly,Whalen2,Schlag,UltracoldChem,RydbergRev,RostLuukko,FeyNew}.

Simultaneously, the production of ultracold mixtures in the quantum degenerate regime has become routine in numerous atomic species \cite{FermiBose1,FermiBose3,KRbmix,KCsmix}. Many experiments can now be performed, for example to study sympathetic cooling of different species, or to study recombination between species leading to unusual few-body bound states such as Efimov trimers \cite{FermiBose2,KRbHetEfimov,LiCsHetEfimov}. Progress in the study of ultracold polar molecules has resulted in the creation of nearly every bi-alkali combination; this is a promising route towards the production of ultracold molecules \cite{Carr,chem1,Ni231,Danzl1062,KremsColdChem,Mols1}. The reactivities, electric and magnetic dipole moments, and other molecular properties vary greatly between bi-alkali pairings. 

This present article has three primary aims motivated by these recent experimental developments. First, to encourage the study of Rydberg molecules composed of Li, Na, and K by calculating their low energy electron scattering phase shifts. These calculations are more specialized to the very low energy scattering regime important for Rydberg molecules than most existing calculations. Second, to systematically display potential energy curves for these Rydberg molecules and show the diversity of molecular features stemming from the broad parameter space of atomic properties, e.g. quantum defects, scattering phase shifts, and hyperfine splittings. Mixing between the polar trilobite-type states and non-polar low-$l$ states occurs frequently in these molecules, providing new experimental opportunites. Third, to show that spectroscopy of these heteronuclear molecules reveals details of the relative densities and spatial distributions of atomic species in a two-component mixture, and thus can be used to determine the properties of quantum degenerate mixtures over a range of length scales via Rydberg spectroscopy.

\section{Scattering phase shifts}
\label{theorysec}
Although atom-electron scattering phase shifts for the light alkali atoms were first calculated more than forty years ago, the extreme sensitivity of Rydberg molecule potential curves, particularly at the kHz-level resolution of current experiments, demands highly accurate phase shifts. Due to the increase in computational power and refined experimental measurements since these earlier calculations, it is valuable to revisit these calculations. Electron-atom scattering phase shifts are calculated using a nearly \textit{ab initio} two-electron model Hamiltonian. The atomic valence electron and the scattering electron interact individually with the positive ion through a model potential $V_l(r)$ \cite{Marinescu} \footnote{Specifically, Eqs. 18a and 18b of [45] give the model potentials. We used the same parameters given in [45]. The independence of our results with respect to the model potential was tested by using the model potential given by Ref. [47] for potassium. }  which depends on the electronic orbital angular momentum $l$.  With these model potentials the Hamiltonian (Atomic units will be used throughout) is
 \be \label{2eham}
 H =\sum_{i=1}^2\left[ -\frac{\nabla_i^2}{2} + V_{l_i}(r_i)\right] + \frac{1}{|\vec r_1 - \vec r_2|} + V_\text{pol}(\vec r_1,\vec r_2),\ee 
where $\vec r_1,\vec r_2$ and $\vec\nabla_1,\vec\nabla_2$ are, respectively, the electron position and momentum operators, $\frac{1}{|\vec r_1 - \vec r_2|}$ is the Coulomb repulsion between the two electrons, and $V_\text{pol}$ is the dielectronic polarization term
\be
V_\text{pol}(\vec r_1,\vec r_2)=-\frac{\alpha_{\text{ion}}}{r_1^2r_2^2}(1 - e^{-(r_1/r_c)^3})(1 - e^{-(r_2/r_c)^3})\cos\theta_{12}.\nonumber
\ee
This models how the electrons interact indirectly through the induced polarization of the ionic core, which has a polarizability $\alpha_\text{ion}$ \cite{OrangeRMP}. The cutoff radius $r_c$ is determined by fitting the electron affinity (EA) to experiment (see Table \ref{tab:phases}). Neglecting $V_\text{pol}$ exaggerates the EA, which led early studies of Cs to erroneously conclude that its $^3P$ shape resonances were weakly bound states \cite{ChrisJJ,pwaveresonance1}.

We use the eigenchannel $R$-matrix method to solve the time-independent Schr\"{o}dinger equation with the Hamiltonian in Eq. \ref{2eham} and standing wave scattering boundary conditions. Since many references extensively discuss this method and its use in such diverse applications as Rydberg photoionization, negative ion photodetachment, and electron scattering \cite{TaranaCurikLi,Sadeghpour2,tennyson,GreeneAymar,EilesNegIon,OrangeRMP}, here the description focuses on the specific numerical implementation. We varied the size of the $R$-matrix region from $25a_0$ to $55a_0$, distances large enough to contain a few low lying wave functions and also to ensure that the long-range potential outside of the $R$-matrix region is dominated by centrifugal and polarization contributions. Although quantum defect theory could be used to directly match to appropriate asymptotic solutions \cite{WatanabeGreene}, we instead numerically propagated the wave function in this long-range potential from the inner $R$-matrix boundary out to a much larger distance $\sim2\times 10^4a_0$, then matched it to the standing wave free-particle solutions to determine the phase shifts \cite{PanStaraceGreene1}. The phase shifts are independent of this outer distance, and down to 200 meV are independent of the size of the $R$-matrix region. At lower energies they depend, at the order of a few percent, on the size of the $R$-matrix region.  Based on these convergence tests we fixed the $R$-matrix region size to be $45a_0$ for K and $35a_0$ for Na and Li. The larger value for K is consistent with its larger polarization. At extremely low energies the phase shifts become unphysical due to the finite $R$-matrix region, and we fit the scattering lengths to the effective range formula of Ref.  \cite{OmalleyRosenbergSpruch}. This fit is insensitive to the range of fitting energies, and allowed for accurate extrapolation to zero energy. We included 10 partial waves in the two-electron basis, constructed out of 48 closed functions and 2 open functions computed with a B-spline basis. This treatment is non-relativistic, so the $^{1,3}S$ and $^{1,3}P$ symmetries were computed separately.

 \begin{figure}[h]
\includegraphics[width= 0.92\columnwidth]{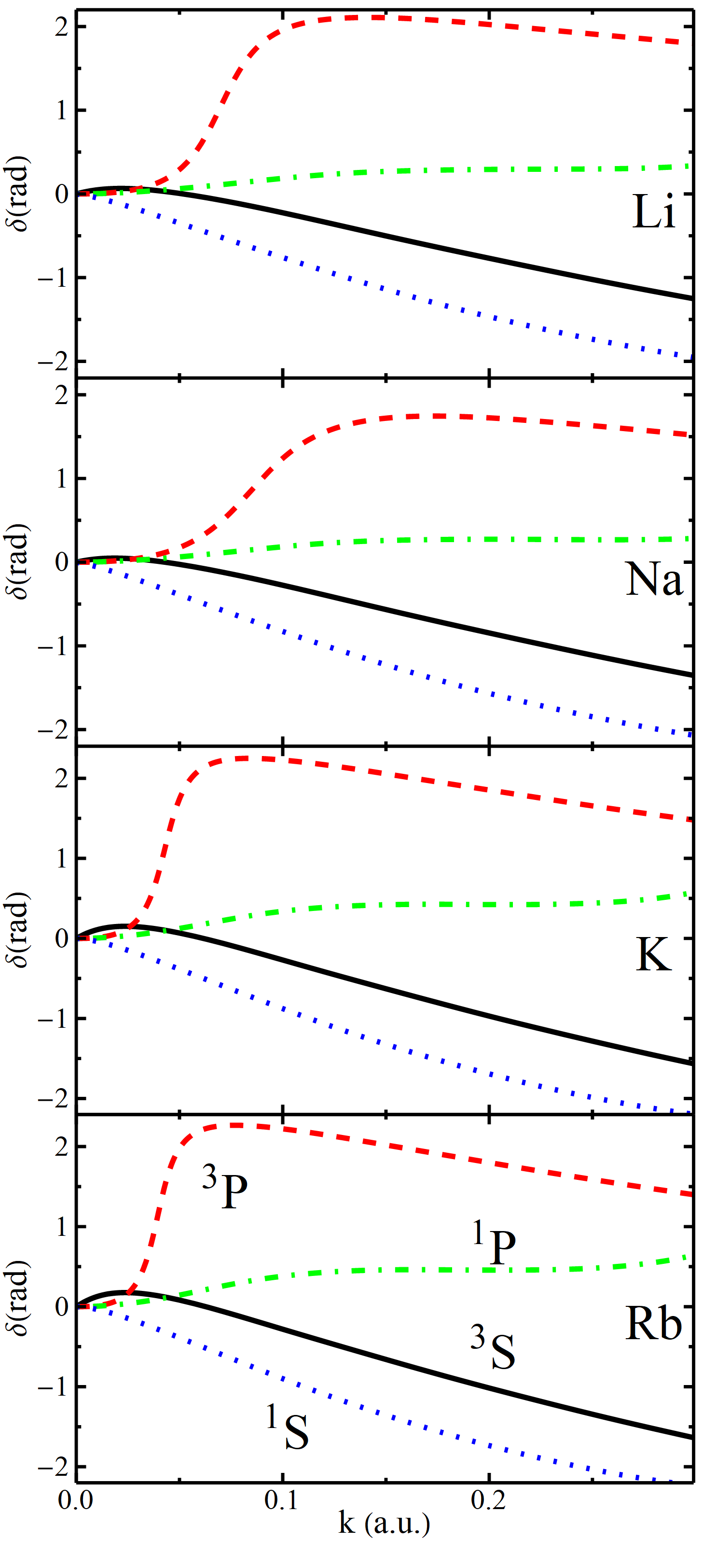}
\caption{\label{fig:phaseshifts} Li, Na, K, and Rb phase shifts. The color scheme corresponds to the labeling in the bottom panel.  The range of momentum values extends nearly to the first excited atomic threshold in every case, covering the entire range of energies relevant for Rydberg molecules. }
\end{figure}

\begin{center}
\begin{table*}[t]
\begin{tabular}{||c|| c| c | c | c | c | c || }
\hline

  &$a_s^T(a_0)$&$a_s^S(a_0)$ &$E_p^T$ (meV) & $\Gamma_p^T$ (meV) & EA$_\text{exp}$ (meV)  & EA$_\text{th}$ (meV)\\ 
  \hline \hline

 Li& $-7.12^a,-7.43^b,-5.66^d,-6.7^\dagger$ & $3.04^a,2.99^b,3.65^d,3.2^\dagger$ & $60.9^b,60^c,63^\dagger$ & $67.9^b,57^c,62^\dagger$ &  618.049 \cite{HaefflerEA} & $618.038^i, 621.77^j$ \\ 
\hline
 Na &  $-6.19^a,-5.9^d,-5.7^\dagger$ & $4.03^a,4.2^d,4.2^\dagger$ & $83^c,80^e$,$83^f,87.9^\dagger$ & $85^c,188^f,124.8^\dagger$ & 547.930 \cite{HotopReview} & $547.539^i$, $563.25^j$ \\ 
\hline
 K &   $-15^d,-15.4^g,-14.6^\dagger$ & $0.55^d,0.57^g,0.63^\dagger$ & $2.4^c,19^g,20^h$ & $.6^c,16^g$ & 501.459 \cite{AnderssonKEA} & $501.231^i, 544.87^j$\\
  \hline
\end{tabular}
\caption{A summary of previous theoretical values and our calculated values (marked by $^\dagger$) for the zero-energy scattering lengths for the triplet (T) and singlet (S) states and the positions and widths of triplet $p$-wave shape resonances. Literatures values are from: a) \cite{LiNaNorcross}, b) \cite{TaranaCurikLi}, c) \cite{Nesbet1973}, d)\cite{Karule}, e)  \cite{JohnstonBurrow}, f) \cite{BFKNa}, g) \cite{Fabrikant1986}, h)  \cite{Moores1976}. Theoretical uncertainties for the literature values are unknown; as discussed in the main text we estimate an uncertainty of $\pm0.2a_0$ on our zero-energy scattering lengths and $\pm 1$meV on the resonance parameters. The final two columns display the experimental electron affinities and those calculated here, with (i) and without (j) the dielectronic polarizability term in the Hamiltonian. 
}
 \label{tab:phases}
\end{table*}
 \end{center}

\begin{figure}[b]
\includegraphics[width= 0.49\columnwidth]{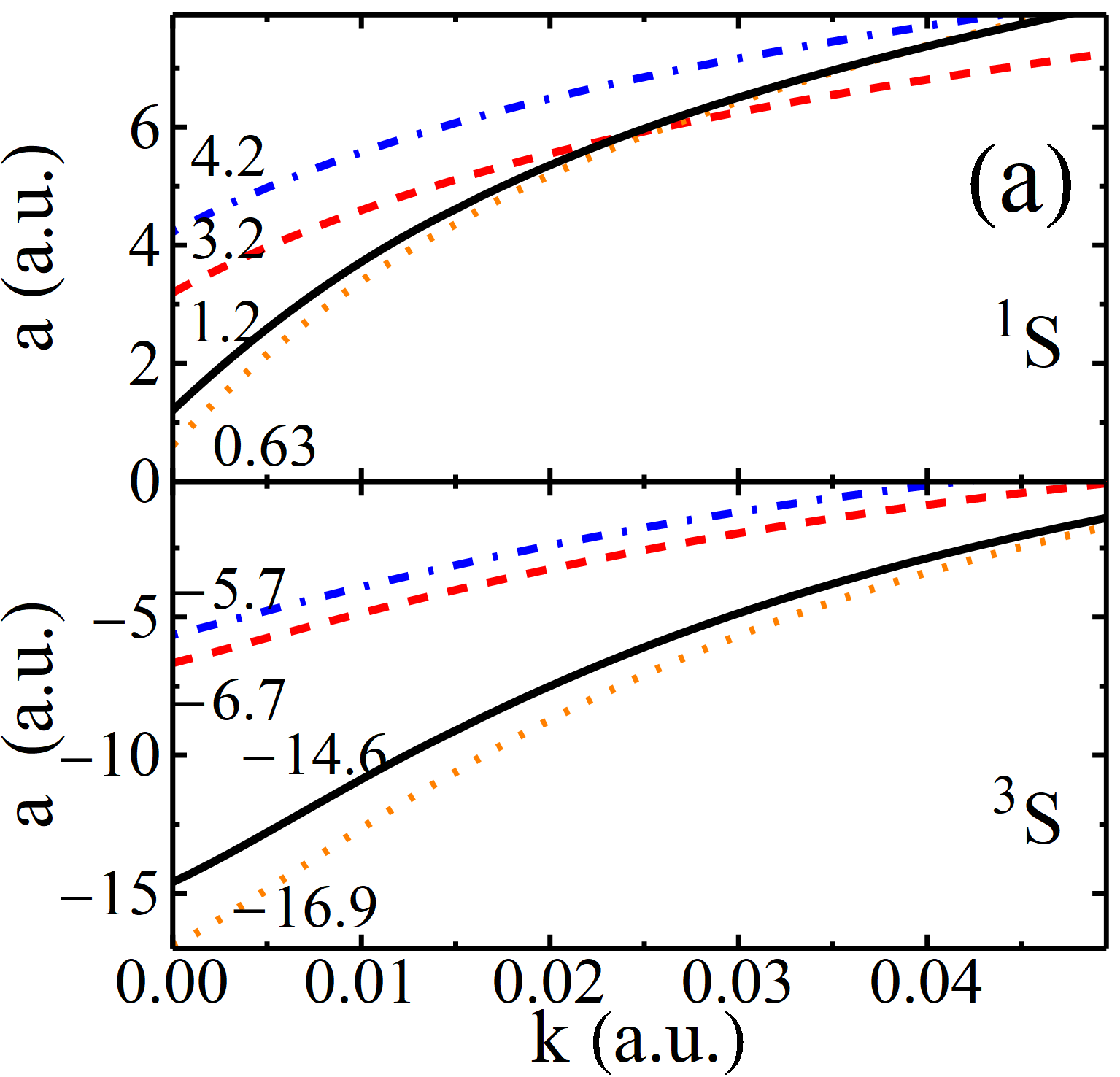}%{D:/PhaseShifts/AllphaseDetails1.png}
\includegraphics[width= 0.49\columnwidth]{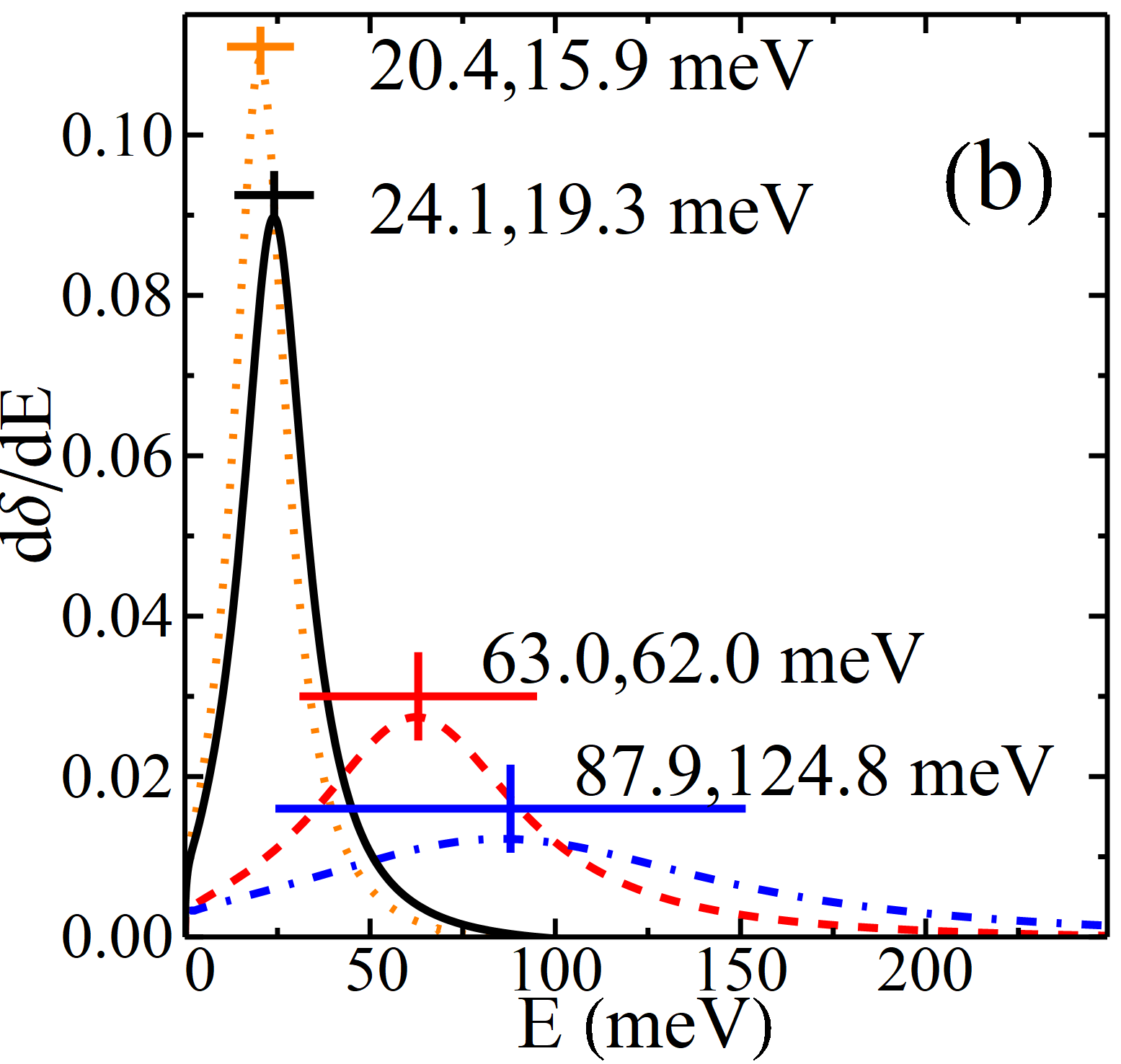}%{D:/PhaseShifts/AllphaseDetails2.png}
\caption{\label{fig:shaperesonances} (a) Singlet and triplet scattering lengths and their zero-energy values for Li (red, dashed), Na (blue, dot-dashed), K (black, solid), and Rb (green, dotted).  (b) Energy derivatives of $^3P$ phase shifts revealing the shape resonances in all three light alkalis. Their positions $E_r$ and widths $\Gamma_r$ are labeled $E_r,\Gamma_r$ for Li (red, dashed), Na (blue, dot-dashed), K (black, solid), and Rb (green, dotted).}
\end{figure}

Fig. \ref{fig:phaseshifts} displays calculated phase shifts for Li, K, Na, and non-relativistic Rb. This latter set of phase shifts is included for comparison with the results of Refs.\cite{RbCsFr,Fabrikant1986} to support the accuracy of the other results. Relativistic Rb and Cs phase shifts have been calculated in Ref. \cite{KhuskivadzePRA}. In all these atomic species the $^3S$ phase shifts  exhibit a low-energy peak, characteristic of a virtual state and therefore a negative scattering length, while the $^1S$ phase shifts all increase monotonically to their zero-energy value and have small positive scattering lengths.  All atoms possess a $^3P$ shape resonance, which is quite broad in Li and especially Na. Fig. \ref{fig:shaperesonances}a shows the $s$-wave scattering lengths and their smooth extrapolation to zero energy. Fig. \ref{fig:shaperesonances}b shows the $^3P$ resonance profiles and their positions and widths obtained fits to Lorentz profiles. Tabulated values of the phase shifts can be found in the supplemental material \footnote{Supplementary material available online}.

Table \ref{tab:phases} summarizes previous results from the literature. Our Li phase shifts agree with those calculated recently in Ref. \cite{TaranaCurikLi}, although the zero-energy scattering lengths differ by around 10\%. The effect of the dielectronic polarizability term in such a small atom is tiny, and had almost no effect on these results; likewise, the calculated polarizability equaled the experimental value (see Tab. \ref{tab:datatable}) with less than one percent error.  It appears that the dominant source of error in the zero-energy values stems from the dependence on the $R$-matrix region size, but it is unclear if this explains the discrepancies with previous results as none of the box sizes  studied here led to zero energy values as large in magnitude as those previously reported. 

The Na results from Ref. \cite{BFKNa} are in good agreement with ours to within a few percent.  The $^3P$ resonance energy is higher than all previous theoretical values, and this appears to be caused by the dielectronic polarizability.  Upon neglecting this term the position of the resonance decreases by approximately 5 meV. The calculated Na polarizability differs slightly from the most recent experimental value, also because of higher order polarization effects neglected in our model potential. For this reason for Na and K we used the experimental polarizability in both the propagation of the wave function outside of the $R$-matrix region and in the effective range expansion to zero energy. Use of the theoretical value slightly depressed the Na triplet scattering length to around -5.8$a_0$, hinting that the discrepancy between the present calculations and the literature values are due in part to their neglect of these dielectronic polarizability induced effects.

\begin{table*}[t]
\begin{tabular}{|| c|| c c| c|| c c| c|| c c||}
\hline
 {\bf Li} & $\mu(0)$ & $\mu'(0)$ &{\bf Na} & $\mu(0)$ & $\mu'(0)$ &{\bf K} & $\mu(0)$ & $\mu'(0)$  \\ 
  \hline \hline
  $s_{1/2}$& 0.3995101 & 0.0290 & $s_{1/2}$& 1.347964 & 0.060673 & $s_{1/2}$& 2.1801985 & 0.13558   \\
 $p_{1/2}$ & 0.0471835(0.0471780) & -0.024 & $p_{1/2}$& 0.855380 & 0.11363  & $p_{1/2}$& 1.713892 & 0.233294 \\
 $p_{3/2}$ & 0.0471720(0.0471665) & -0.024 & $p_{3/2}$& 0.854565 & 0.114195 &$p_{3/2}$& 1.710848 & 0.235437 \\
 $d_{3/2}$ & 0.002129 & -0.01491 & $d_{3/2}$& 0.015543 & -0.08535 & $d_{3/2}$& 0.2769700 & -1.024911  \\
 $d_{5/2}$ & 0.002129 & -0.01491& $d_{5/2}$& 0.015543 & -0.08535 & $d_{5/2}$& 0.2771580 & -1.025635 \\
 $f_{5/2}$ & -0.000077 &0.021856 & $f_{5/2}$& 0.0001453 & 0.017312 & $f_{5/2}$& 0.010098 & -0.100224   \\
 $f_{7/2}$ & -0.000077 &0.021856 & $f_{7/2}$& 0.0001453 & 0.017312 & $f_{7/2}$& 0.010098 & -0.100224     \\
 \hline \hline
  & $\alpha$ (a.u.) & $\alpha_\text{ion}$ (a.u.) &   & $\alpha$ (a.u.)  & $\alpha_\text{ion}$ (a.u.)  & & $\alpha$ (a.u.)  & $\alpha_\text{ion}$ (a.u.) \\
  \hline
  {\bf Li} & 164.9$^{a,b}$ & 0.1923 & {\bf Na}  & 165.9$^a$,162.7$^b$ & 0.9448 & {\bf K} & 307.5$^a$,290.6$^b$ & 5.3310\\
  \hline \hline
   & $^{6}$Li $(i=1)$ & $^7$Li $(i=3/2)$ &  & $^{23}$Na&    &  & $^{39}$K& $^{41}$K\\
   \hline
   A$_\text{Li}$ (MHz) & $152.137$ & 401.752  & A$_\text{Na}$ (MHz) & 885.813&   &  A$_\text{K}$ (MHz) & 230.86 & 127.01 \\
 \hline
 %Li: 228.2052590,  803.5040866. Na: 1771.6261288, K: 461.719720,  254.013872
\end{tabular}
\caption{Quantum defects, atomic polarizabilities $\alpha$, ionic polarizability $\alpha_i$, and hyperfine constants $A$ for the most common light alkali isotopes $^{6,7}$Li, $^{23}$Na, and $^{39,41}$K. The hyperfine constants are from \cite{Beckmann1974}, the Li quantum defects are from \cite{GoyLi} and \cite{NiemaxLi}, and Na and K quantum defects are from \cite{NiemaxLi}.  The value for $p_J$ in parentheses is for $^7$Li, while the rest are isotope-independent. $^a$ The polarizabilities calculated using the model potential in our $R$-matrix calculations; $^b$ polarizabilities measured in Ref. \cite{updatedPols} and \cite{Miffre2006}. A similar table containing values for Rb and Cs can be found in Ref. \cite{EilesSpin}.
}.

 \label{tab:datatable}
\end{table*}

The K phase shifts of Ref. \cite{KaruleOld} are, to the best of our knowledge, the only detailed calculation of these phase shifts over this energy range, although Refs. \cite{Fabrikant1986,Moores1976} report scattering lengths and resonance parameters. In K the dielectronic polarizability plays a much greater role than it does in Li and Na. Our $^1S$ scattering length is significantly larger than previous values; however, by neglecting the dielectronic polarization term in the Hamiltonian or using the theoretical rather than experimental polarizability for the long-range potential we obtain scattering lengths ranging from 0.2 - 0.9 $a_0$. Likewise, the $^3S$ scattering length changes to -15.1$a_0$ when the theoretical polarizability is used. The $^3P$ shape resonance position shifts to much lower energies without the dielectronic polarization term.

Overall, the present results are generally in good agreement with previous calculations over a large range of energies. The discussion above suggests that significant differences are likely caused by the inclusion of higher order polarization effects in the model Hamiltonian via the dielectronic polarization term and the use of accurate empirical dipole polarizabilities.  Sources of error stemming from the model potential, finite $R$-matrix region, prematurely truncated two-electron basis set, or errors in the extrapolation to zero energy, all contribute to an uncertainty in these quoted scattering lengths and resonance parameters which we estimate to be $\pm 0.2a_0$ and $\pm 1$meV, respectively.  The good agreement between the Rb results and previous non-relativistic calculations further supports these conclusions \cite{RbCsFr,Fabrikant1986}.

\section{Ultra-long-range Rydberg molecules}
\label{dimers}

 \begin{figure*}[t]
\includegraphics[width= \textwidth]{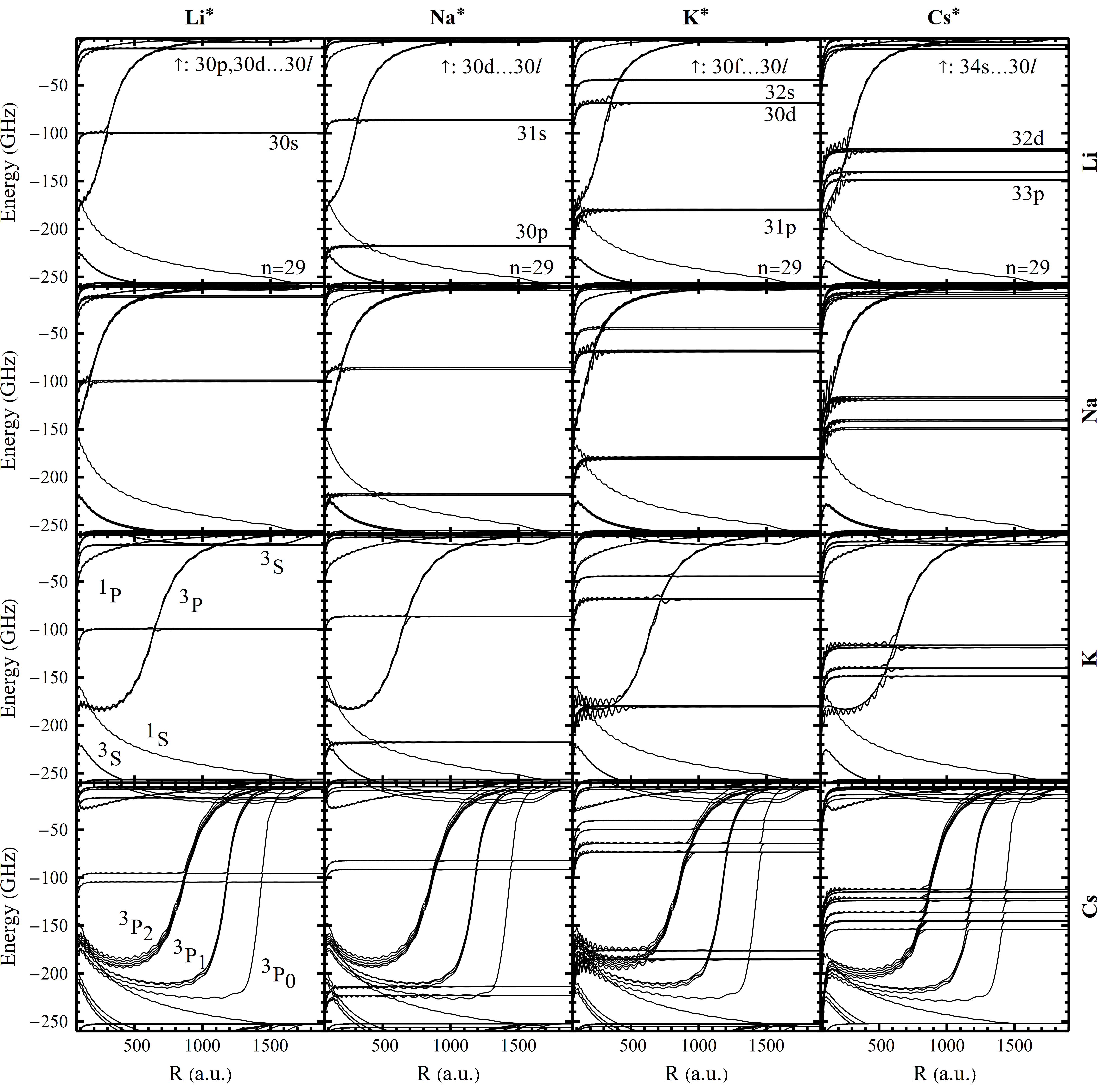}
\caption{\label{fig:fullspectrum} Diatomic potential energy curves for X$^*$Y systems with $\Omega = 1/2$. Each column is a different Rydberg atom X$^*$, and each row a different ground state atom Y. Homonuclear dimers are thus along the diagonal. The potential curves can be labeled as $R\to\infty$ by the isolated Rydberg atom's quantum numbers $n,l$ as is done in the top row. Nearly degenerate levels lying very near the hydrogenic manifold are written following the $\uparrow$ label. The dominant potential curves for different scattering symmetries are labeled in the Li$^*$K and Li$^*$Cs panels. The energy scale is relative to the $n=30$ hydrogen energy.  }
\end{figure*}

\begin{figure*}[t]
\includegraphics[width= \textwidth]{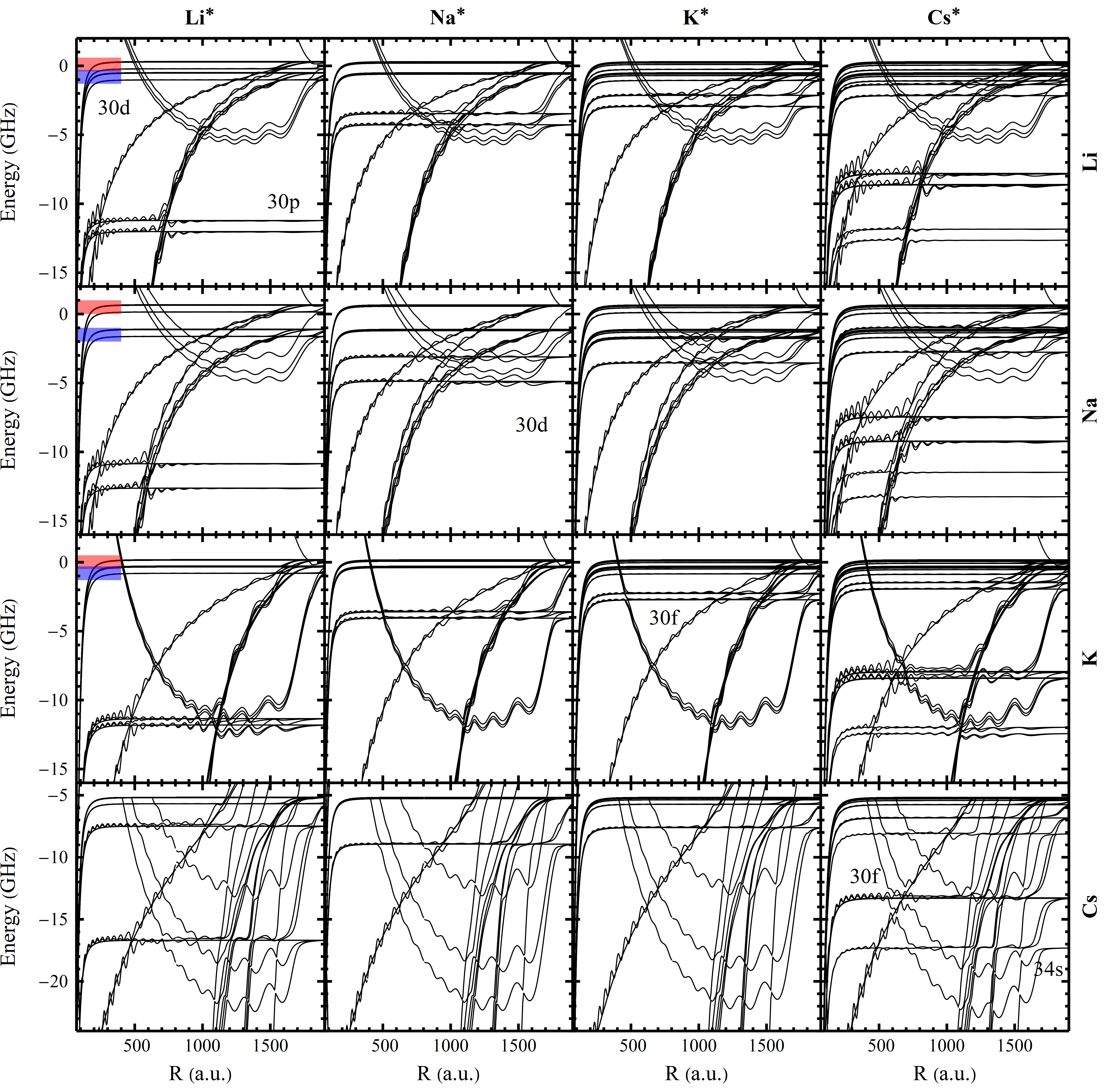}
\caption{\label{fig:trilobites} The same as Fig. \ref{fig:fullspectrum}, but showing just the trilobite region near the $n=30$ degenerate manifold. The degenerate manifolds, split by the perturber's hyperfine splitting, are marked in the first column by red ($F = 2$) and blue ($F = 1$), where $F$ is the total spin of the perturber. Cs, in the bottom row, has instead $F = 3,4$ and for clarity only the $F = 3$ hyperfine manifold is shown. The $nl$ labels are refer to the nearest low-$l$ potential curve and correspond to the quantum numbers of the isolated Rydberg atom.  The energy scale is relative to the $n=30$ hydrogen energy. }
\end{figure*}

The $s$ and $p$-wave phase shifts summarized in the previous section, along with the empirical quantum defects, polarizabilities, and hyperfine constants given in Table \ref{tab:datatable}, provide all the numerical and empirical input required to calculate Rydberg molecule potential energy curves. This calculation is outlined in detail in Ref. \cite{EilesSpin}, and so we only briefly summarize the main ideas here. The total Hamiltonian includes the individual Hamiltonians of the non-interacting Rydberg and ground state atoms along with the spin-dependent interaction potential described by the $s$-wave Fermi pseudopotential and $p$-wave generalization due to Omont \cite{Fermi,Omont}. The energies of the isolated Rydberg atom are determined by their quantum defect parameters $\mu(0)$ and $\mu'(0)$ as defined, e.g., in Ref. \cite{EilesSpin}. The full electronic Hamiltonian for the Rydberg molecule is diagonalized in a basis of Rydberg states with total angular momentum $j$ coupled to the nuclear and electronic spins $i$ and $s_2$ of the perturber. The eigenenergies of this Hamiltonian, evaluated at each internuclear separation $R$, are the adiabatic potential energy curves which determine the rovibrational structure of the Rydberg molecule. We also include the polarization potential $-\frac{\alpha}{2R^4}$ between the Rydberg ion and the ground state atom. Since the light alkali atoms have small spin-orbit splittings scaling as $Z^4$ we have only used a non-relativistic $R$-matrix calculation to compute phase shifts, and therefore do not include $J$-dependent phase shifts as is necessary for Rb and Cs. This approximation is excellent for Li and Na, and is sufficiently valid for K, although its similarity to Rb suggests that the spin-orbit splitting of the $p$-wave phase shifts could affect butterfly-like states. 

Like Rydberg atoms, Rydberg molecules follow several $n$-scaling laws. The energy splittings between Rydberg states decrease as $n^{-3}$. The molecular potential wells determined by $s$-wave scattering decrease as $n^{-6}$ for the  low-$l$ states, while the trilobite potential wells, formed by the mixing of degenerate high-$l$ states, decrease as $n^{-3}$. In contrast, the effect of the $p$-wave shape resonance is approximately independent of $n$, and the hyperfine splitting of the ground state atom is fully independent of $n$. Thus, only some of the aspects of these potential curves investigated here scale rigorously with $n$. The different scaling laws can lead to a variety of new phenomena. Examples can be found in Refs. \cite{Niederprum,PfauRaithel} where the mismatch between the hyperfine, Rydberg, and fine structure energy scales was utilized to exploit favorable degeneracies. Ref. \cite{BoothTrilobite}, on the other hand, utilized near-degeneracies in the $s$ quantum defect that are independent of $n$ to admix trilobite and $ns$ states. 

To match previous theoretical work \cite{EilesSpin,Greene2000,KhuskivadzePRA} we set $n=30$ as the prototype Rydberg state. As discussed in \cite{EilesSpin,Fey}, there are considerable ambiguities associated with the rigorous convergence of this calculation due to singularities in the Fermi pseudopotential and the divergent $p$-wave phase shift. For consistency we have, as in Ref. \cite{EilesSpin}, chosen a basis of Rydberg states with $n\in \{28,\dots,31\}$, and have set the semiclassical momentum to $[k(R)]^2=2(-\frac{1}{2(30)^2} +\frac{1}{R})$.  The only good quantum number of the system is the projection of the total angular momentum onto the internuclear axis, $\Omega$, and we focus on $\Omega = \frac{1}{2}$ for all dimers since this symmetry permits the most potential curves. At infinitely large internuclear separations the Rydberg atom's angular momentum $\vec j = \vec l + \vec s_1$ and total spin of the ground state atom $\vec F = \vec s_2 + \vec i$ are good quantum numbers and are used to label the energy levels reached asymptotically by each potential energy curve. The Rydberg molecule potential curves can be approximately labeled by the dominant scattering channel quantum numbers, $^{2S+1}L$, where $\vec S = \vec s_1 + \vec s_2$ and $\vec L$ is the orbital angular momentum of the Rydberg electron relative to the perturber. These are typically called singlet(triplet) trilobite(butterfly) potentials, for $S=0(1)$ and $L = 0(1)$, respectively.

Fig. \ref{fig:fullspectrum} displays these potential curves for all heteronuclear combinations X$^*$Y, where X$^*$ denotes the Rydberg atom and Y denotes the ground state atom. Fig. \ref{fig:trilobites} highlights the trilobite potential wells. In addition to Li, Na, and K we consider Cs, the heaviest alkali atom regularly studied in the laboratory, because of the interesting physics associated with its large nuclear spin ($i=7/2$) and pronounced $^3P_J$ splittings. It is also a promising candidate for heteronuclear systems involving K \cite{KCsmix}. Rb, although very common in ultracold systems, has very similar scattering properties and quantum defects as K, and so for brevity we did not study combinations involving this atom.  We also focus on the most common bosonic isotopes, $^7$Li, $^{23}$Na, $^{39}$K, and $^{133}$Cs, since the primary effect of different isotopes on the potential energy curves presented here is the fairly trivial change in hyperfine splitting. The choice of isotope can, of course, dramatically affect the spatial correlations and stability of quantum degenerate mixtures due the different statistics obeyed by bosonic or fermionic isotopes.

Fig. \ref{fig:fullspectrum} shows the general energy landscape for each dimer combination considered here. The positions of low$-l$ states shifted by their quantum defects are nearly identical within a single column (their positions depend slightly on the row due to the hyperfine splitting of the perturber) and are labeled in the top row.  At this level the fine structure splittings of Li, Na, and K are unresolvable. In Fig. \ref{fig:trilobites} some energy levels in each column are labeled to orient the reader, and the hyperfine-split degenerate manifolds are highlighted to show the size of the hyperfine splitting on this level and distinguish it from fine-structure splittings. To organize the detailed analysis of these potential curves we will consider each atom separately in the following, although these discussions will still overlap somewhat when considering heteronuclear molecules. 
 \begin{figure}[t]
 \begin{center}
\includegraphics[width= \columnwidth]{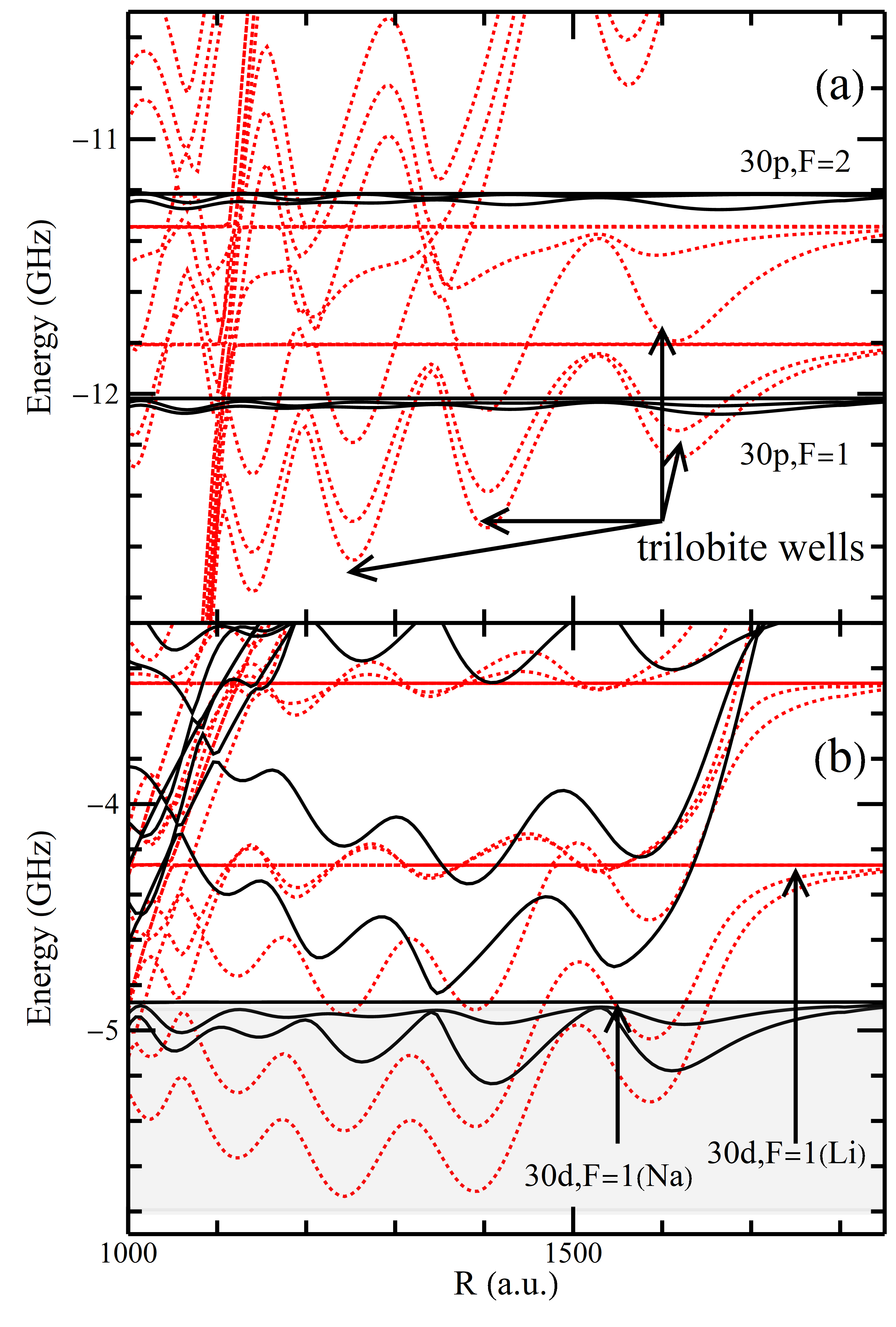}
\caption{\label{fig:TriloMixing} Top: Li$^*$K (dashed red) and Li$^*$Li (black) potential energy curves with $\Omega = 1/2$. Bottom: Na$^*$Li (dashed red) and Na$^*$Na (black) potential energy curves. The gray energy region is discussed further in Sec. \ref{sec:discussion}.  The energy scale is relative to the $n=30$ hydrogen energy. }
\end{center}
\end{figure}

\textbf{\textit{Lithium:  }}All Li$^*$X molecules have relatively featureless potential energy landscapes due to the small Li $l\ne 0$ quantum defects and negligible fine structure splittings. The $np$ state lies close to the hydrogenic manifold and so perturbers with sufficiently large scattering lengths could form trilobites that couple to this state. This is seen in the Li$^*$K and Li$^*$Cs potential curves in Fig. \ref{fig:trilobites} and in Fig. \ref{fig:TriloMixing}a, where the effect of trilobite mixing is evinced by the increased depth of Li$^*$K potentials compared with the Li$^*$Li potentials. The deeper trilobite state of K repels the $np$ Li potentials far more than the weaker Li trilobite state. However, even without this inherent mixing caused by a large scattering length,  external electric fields can mix the $np$ and trilobite state \cite{KurzSchmelcherPRA}. For all X$^*$Li molecules, the small Li-$e$ scattering lengths lead to shallow potentials, almost invisible on the scale of Fig. \ref{fig:fullspectrum}, and its $p$-wave shape resonance is too high in energy to support any butterfly-type states. 
 \begin{figure}[t]
 \begin{center}
\includegraphics[width= \columnwidth]{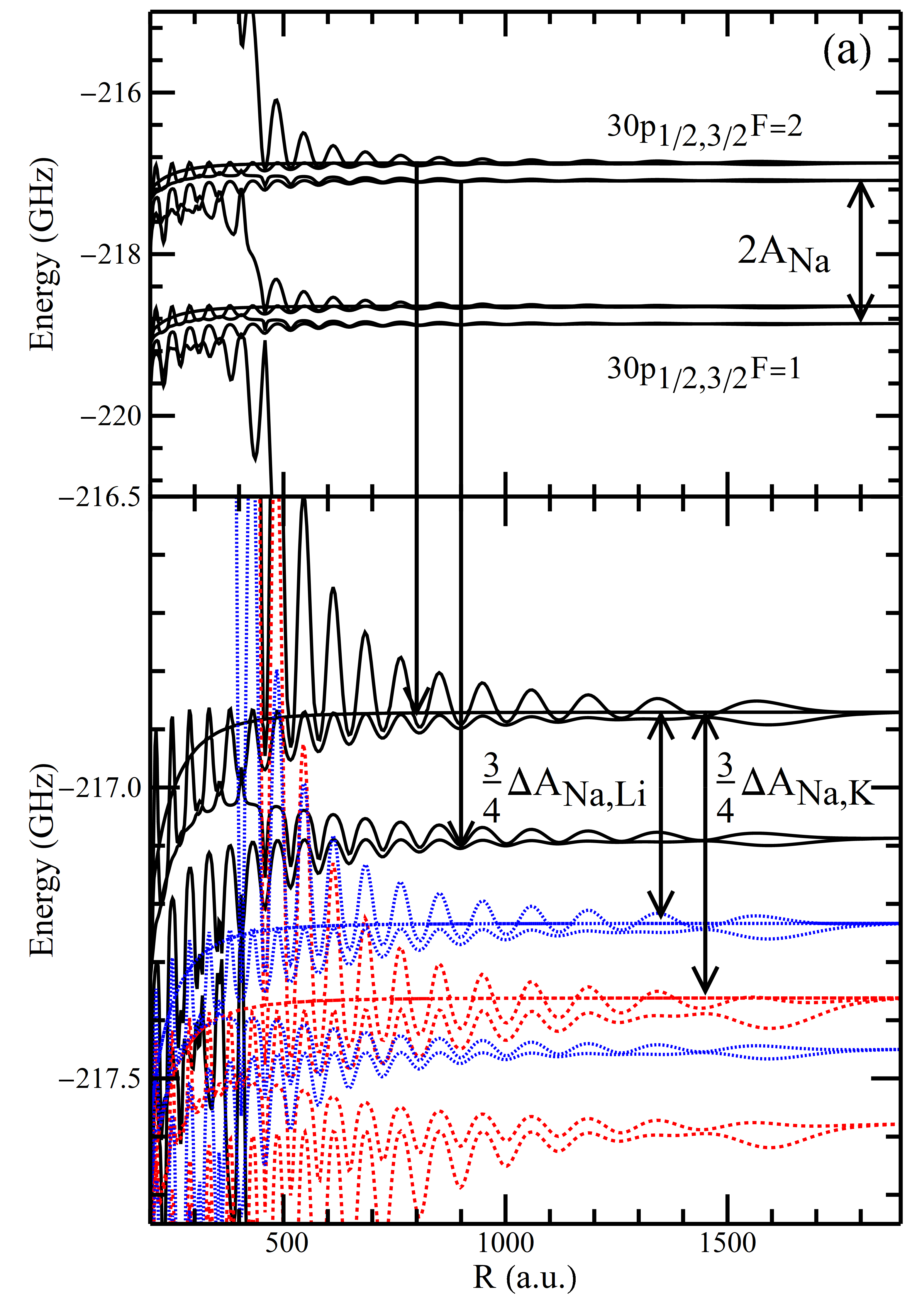}
\caption{\label{fig:pstatesNa} Potential energy curves with $\Omega = 1/2$ associated with the Na $30p_j$ Rydberg states. In (a) Na is the perturber. In (b) the Na$(30p_j)$Na$(F=1)$ (black), Na$(30p_j)$K$(F=1)$ (Red, coarse dashing), and Na$(30p_j)$Li$(F=1)$ (blue, fine dashing) potential curves are shown in greater detail.  The energy scale is relative to the $n=30$ hydrogen energy.  }
\end{center}
\end{figure}

\textbf{\textit{Sodium:  }}All Na$^*$X molecules are characterized by sizeable $s$ and $p$ and very small $d$ quantum defects; these lead to a variety of novel molecular scenarios. The $nd$ state readily mixes with the trilobite state (Fig. \ref{fig:TriloMixing}b). Like the mixed $(n+4)s$ and $n$-trilobite states of Cs \cite{BoothTrilobite}, this creates a pathway for two-photon excitation of highly polar molecules. Unlike in Cs, the $p$-wave contribution and hyperfine splittings in Li and Na are quite weak, so this system is theoretically easier to interpret and experimentally should be more stable and longer-lived. Fig. \ref{fig:TriloMixing}b also shows the different asymptotic energies associate with the hyperfine structure of different perturbers; the energy splitting when both atoms are in the $F=1$ spin state is given by $\frac{5}{4}\Delta A_{\text{Li},\text{Na}}$, where $\Delta A_\text{X,Y} = A_\text{X} - A_\text{Y}$ is the difference between atomic hyperfine constants.

The Na $np$ state is unique among the alkalis in that its energy is close to the $(n-1)$ manifold.  This presents opportunities to excite the repulsive singlet trilobite state, which is also most prominent in Na since it has the largest singlet scattering length, via one-photon excitation through its $p$-state admixture at relatively short internuclear distances (around 500 $a_0$ as highlighted in Fig. \ref{fig:pstatesNa}a). Fig. \ref{fig:pstatesNa}b shows potential curves for Na$^*$Na, Na$^*$K, and Na$^*$Li molecules. In all three cases the effects of the $p$-wave coupling is noticeably absent because the $np$ Rydberg state lies below the butterfly potential wells (Fig. \ref{fig:fullspectrum}), although for both Li and Na perturbers the $p$-wave interaction is too weak to even form these wells. The energy shifts between asymptotic energies are again determined by the difference in hyperfine coupling constants, although now with a different coefficient since these are $F = 2$ levels.  Further insight into the implications of the heteronuclear structure shown in Figs. \ref{fig:pstatesNa}b and \ref{fig:TriloMixing} is given in Sec. \ref{sec:discussion}.

 \begin{figure}[t]
 \begin{center}
\includegraphics[width= \columnwidth]{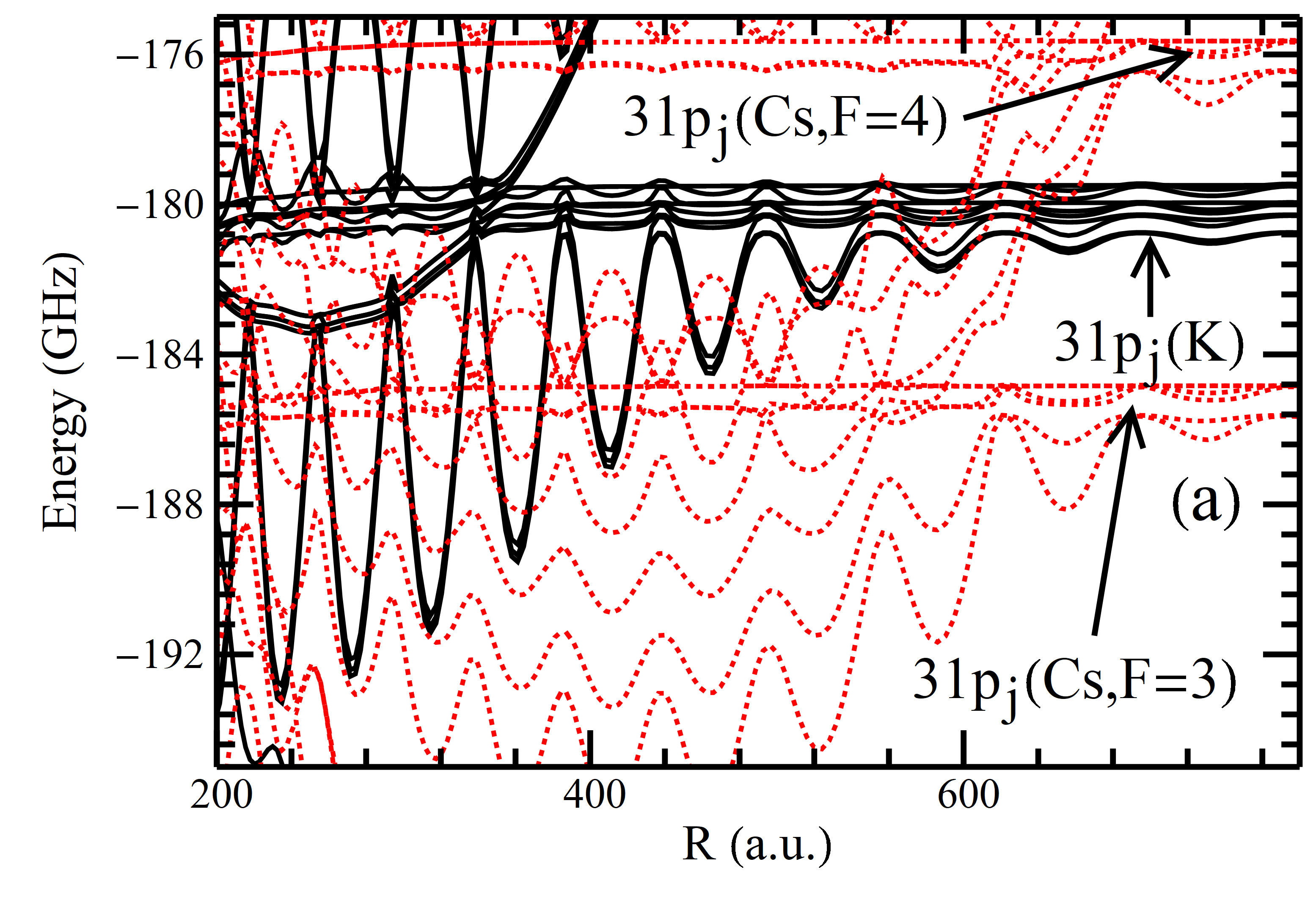}
\includegraphics[width= \columnwidth]{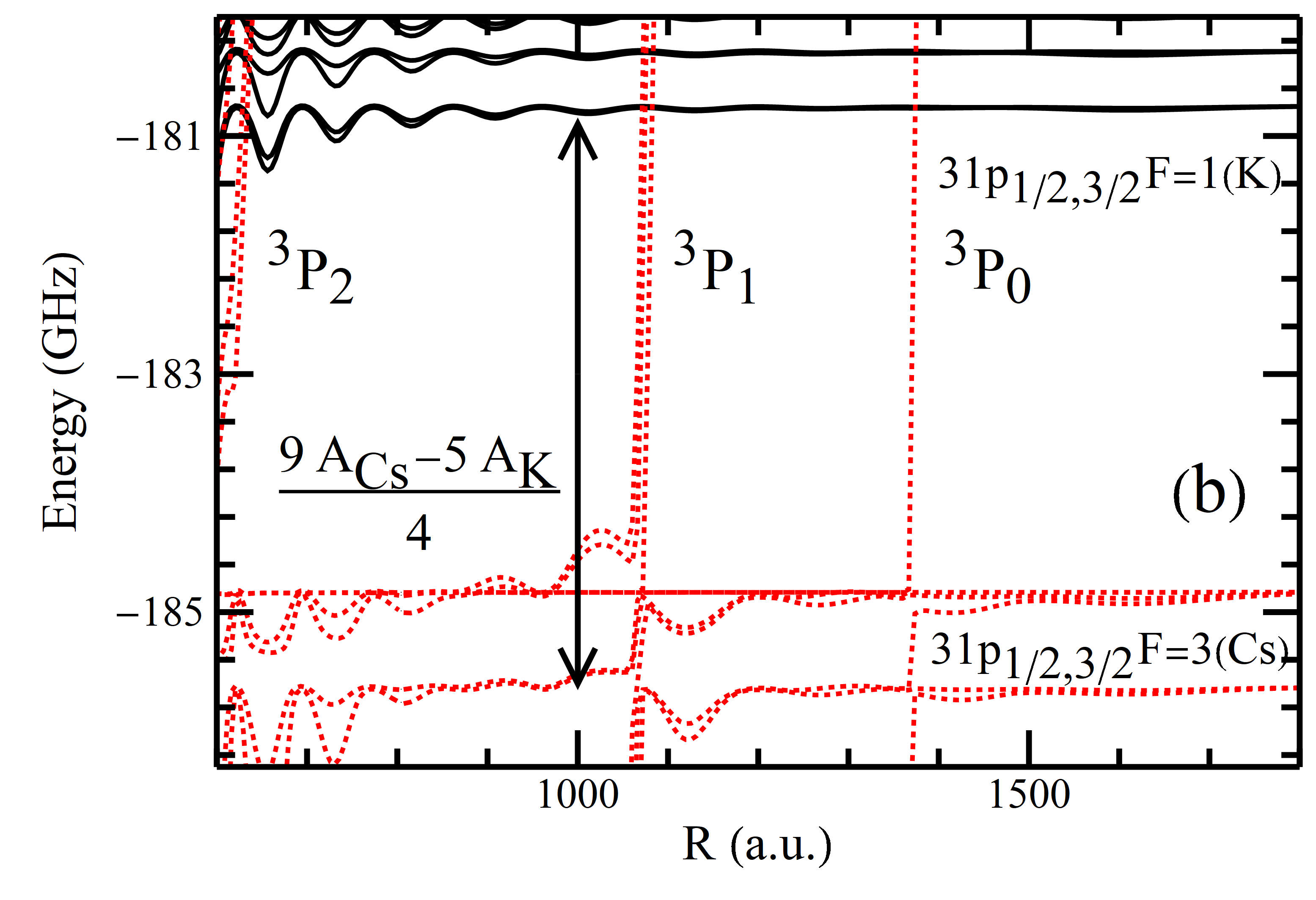}
\caption{\label{fig:cskoverall} Potential energy curves with $\Omega = 1/2$ for K$^*$K (black) and K$^*$-Cs (dashed red) molecules. In (a) the entire region of butterfly states is shown. K$^*$K has several deep, regularly spaced butterfly wells, while K$^*$Cs has many different butterfly potential energy curves since its $^3P_J$ and hyperfine splittings remove many near-degeneracies in the K$^*$K molecule.  The energy scale is relative to the $n=30$ hydrogen energy.  }
\end{center}
\end{figure}

\textbf{\textit{Potassium:  }}Both as the Rydberg atom and as the perturber, K is quite similar to Rb and thus does not have many new features of note that have not been studied extensively in prior investigations. Like the butterfly states observed in Rb \cite{Butterfly}, both K$^*$K and K$^*$Cs butterfly molecules are bound near the $(n+1)p$ state. Fig. \ref{fig:cskoverall}a shows the butterfly potential wells for both of these molecules. Fig. \ref{fig:cskoverall}b shows the $31p_{1/2,3/2}$ Rydberg molecules in greater detail, emphasizing the role of the $p$-wave shape resonances dominant in Cs and again showing how the different hyperfine coupling (now no longer proportional to the difference in $A$ between the two atoms since they have different $F$ values) of the perturber atom leads to overall energy shifts in the asymptotic energies of the atoms.

 \begin{figure}[t]
 \begin{center}
\includegraphics[width= \columnwidth]{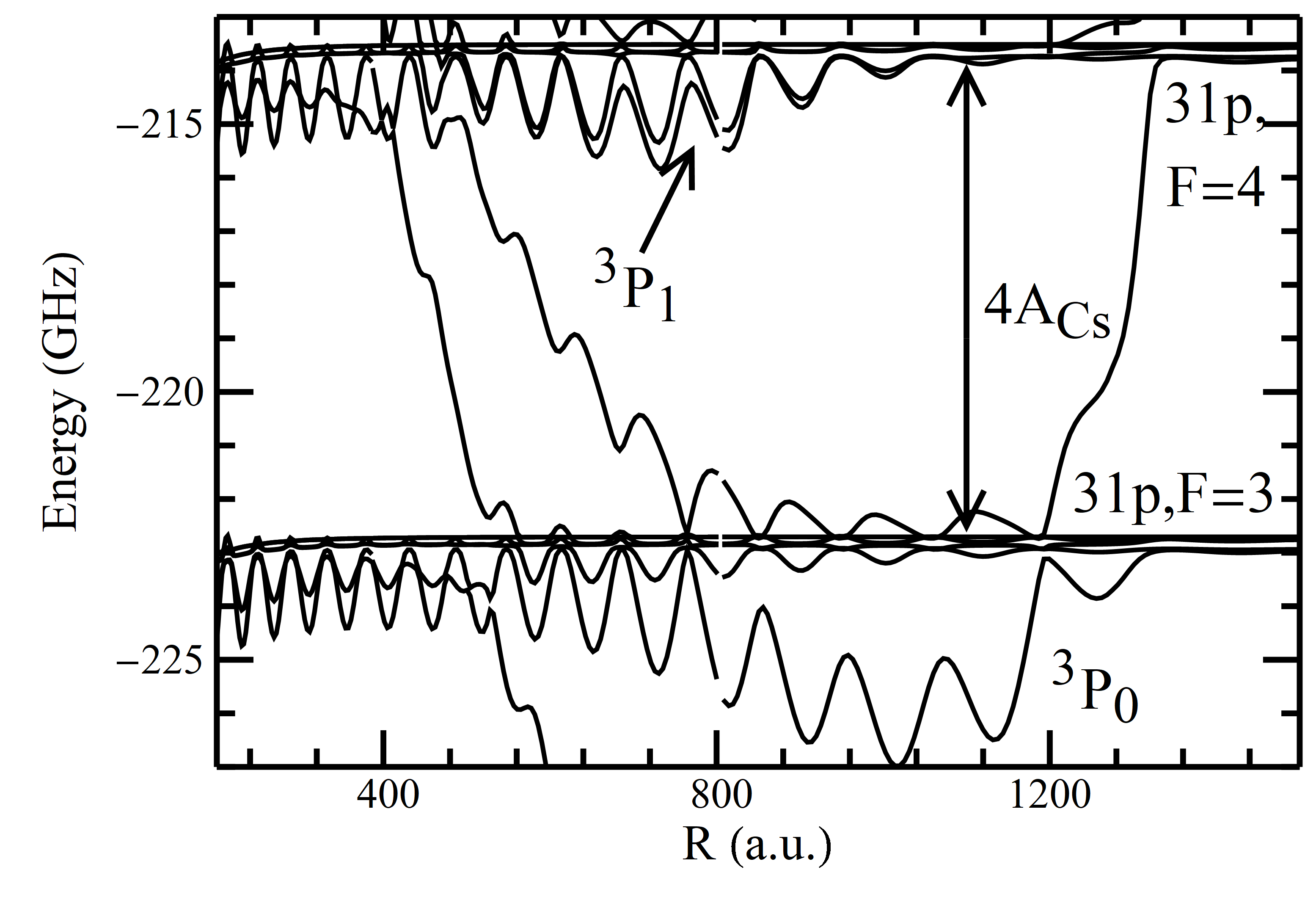}
\caption{\label{fig:butterflyfig} Potential energy curves with $\Omega = 1/2$ supporting unusual long-range butterfly states of the Na$^*$Cs molecule.  The energy scale is relative to the $n=30$ hydrogen energy. }
\end{center}
\end{figure}

\textbf{\textit{Cesium:  }} We have used relativistic Cs phase shifts from Refs. \cite{KhuskivadzePRA,BahrimThumm} in these calculations. Since Cs has been studied extensively before, we only highlight here a particularly interesting set of hybridized butterfly states again caused by the location of the $np$ level of Na. Fig. \ref{fig:butterflyfig} shows that this state cuts through the very long-range butterfly potential wells associated with the $^3P_0$ and $^3P_1$ scattering symmetries. In homonuclear Cs$^*$Cs these states are challenging to excite since they have predominantly high-$l$ electronic character, but in Cs$^*$Na single-photon access should be possible through the admixture of $np$ character to these butterfly states. These are located at much larger internuclear separations than possible in Rb or K, and as a result these butterfly-type molecules will have very large dipole moments and decreased decay rates compared to the butterfly states of Rb \cite{Butterfly}.

\section{Probing ultracold mixtures through Rydberg molecule spectroscopy}
\label{sec:discussion}

The previous section highlighted multiple variations of the basic structure of Rydberg molecules known from previous studies in Rb and Cs. These variations stem from the properties of the light alkali atoms and from the many permutations available when heteronuclear molecules are considered, and lead to many modified molecular states of fundamental interest. Many of these involved hybrid molecules mixing the unusually polar trilobite-like states and the isolated low-$l$ states accessible via typical absorption spectroscopy. These low-$l$ states additionally provide an experimental probe of electron-atom scattering through the dependence of their binding energies on the scattering lengths and $p$-wave resonance positions. The ability to collate data from multiple atomic species, isotopes, and Rydberg states will help to refine these measurements, and show that Rydberg molecules can be used as indirect probes of many atomic properties of fundamental interest.

 \begin{figure}[t]
 \begin{center}
\includegraphics[width=\columnwidth]{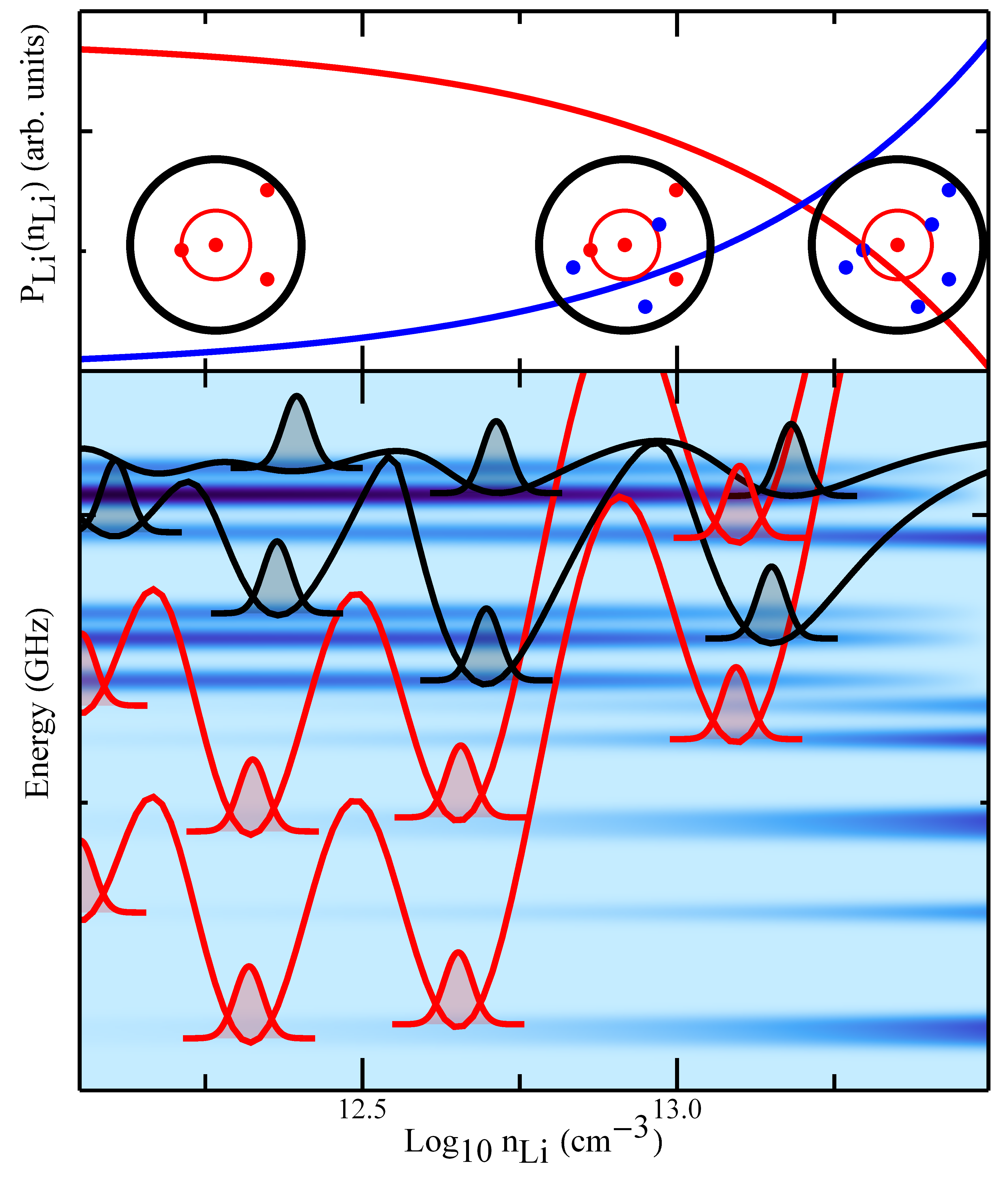}%{D:/PhaseShifts/spectrosigHF.png}
\caption{\label{fig:thepoint} Top: the probabilities $P_\text{Li}(1600a_0)$ (red) and $P_\text{Na}(1600a_0)$ (black) as a function of $n_\text{Li}$ for a fixed total density $n_\text{Li}+n_\text{Na} = 10^{13.5}$cm$^{-3}$.  At $n_\text{Li}\approx 10^{12}$cm$^{-3}$, the gas is effectively single-component and only Na$^*$Na forms. At $n_\text{Li} = n_\text{Na} = 10^{13.2}$cm$^{-3}$ the probability of finding a Li perturber matches that of finding a Na perturber at the desired bond length. Above this density the Rydberg atom is most likely different from the perturber, and heteronuclear molecules predominantly form. Bottom: predicted Rydberg molecule signal with a Na Rydberg atom as a function of $n_\text{Li}$. The homonuclear (black) and heteronuclear (red) potential energy curves are overlayed to show the origin of the spectral lines, which shift from homonuclear to heteronuclear as $n_\text{Li}$ increases. The signal is zero in the light blue regions and strongest when the color is darkest.}
\end{center}
\end{figure}

We now show that Rydberg molecule spectroscopy can also probe the properties of ultracold mixtures, such as the relative densities of different components and their spatial correlations, in a very controllable way.  As Figs. \ref{fig:TriloMixing}, \ref{fig:pstatesNa}, and \ref{fig:cskoverall} suggest, the relative energies of Rydberg molecules to a given atomic line depend strongly on the perturber atom, as its hyperfine structure and well depths can vary drastically. We illustrate this idea with a simple model general to any combination of atoms given here, but for concreteness we will apply it to the Na$^*$Na and Na$^*$Li potential energy curves shown in the highlighted region of Fig. \ref{fig:TriloMixing}. The probability of finding a perturber of species X a distance $r$ from the Na perturber is given by a two-species nearest neighbor distribution,
\be
\label{eq:nnd}
P_\text{X}(r) = \frac{3}{a_\text{X}}\left(\frac{r}{a_\text{X}}\right)^2\exp\left[-\left(\frac{1}{a_{\text{Li}}^3} + \frac{1}{a_{\text{Na}}^3}\right)r^3\right],
\ee
 where $a_\text{X} = \left(\frac{4\pi n_\text{X}}{3}\right)^{-3}$ is the Wigner-Seitz radius for species X, which has a density $n_\text{X}$. Eq. \ref{eq:nnd} can be derived following the same logic as that underlying the nearest neighbor distribution in an ideal gas \cite{Chandra}. This formula is valid for constant density in a three-dimensional gas and neglects the effects of a trapping potential on the actual density distribution of atoms. The results in this section will therefore change depending on the dimensionality and actual density profile of different trapped mixtures. $P_\text{Li}$ and $P_\text{Na}$ are shown as functions of $n_\text{Li}$ in Fig. \ref{fig:thepoint} for a fixed distance $r = 1600a_0$ where the deepest bound molecular states in the low-$l$ potential curves of Fig.\ref{fig:pstatesNa} are localized.  To witness how these relative densities can be measured experimentally, we consider the trilobite-like states of Na$^*$Na and Na$^*$Li shown in the gray region of Fig. \ref{fig:TriloMixing} and reproduced in Fig. \ref{fig:thepoint}.  As a crude approximation of the full vibrational structure of these molecular states, we assume that each deep potential well supports a bound state at its minimum, thus obtaining four binding energies $E_i^\text{Na}$ and bond lengths $R_i^\text{Na}$ for the homonuclear molecule and eight $E_i^\text{Li},R_i^\text{Li}$ for the heteronuclear molecule.  We assume that the Franck-Condon factors for these states are equal. This is reasonable at ultracold temperatures since the initial scattering state of the nuclei is nearly flat over these distances and the transition dipole moments for these different molecular states are very similar. Thus, the photoassociation line strengths are proportional to the probability of finding a Na or K perturber at the right position. With these assumptions, a simple expression for the line strength as a function of energy is given by summing over the line profiles $f(E_i^X)$, centered at $E_i$, and weighted by the probability of finding a perturber at the corresponding $R_i$:
  \be
 S(E) \propto \sum_{X\in \{\text{Na},\text{Li}\}}\sum_iP_X(R_i^X)f(E_i^X),
 \ee
The bottom panel of Fig. \ref{fig:thepoint} shows this spectrum $S(E)$ as a function of $n_\text{Li}$, still with the total density fixed at $n = 10^{13.5}$cm$^{-3}$. At very low $n_\text{Li}$ all that is visible is the homonuclear signal. The lines at a constant density vary in strength because of their different bond lengths. As $n_\text{Li}$ increases new lines become visible as heteronuclear molecules start to become probable. At high $n_\text{Li}$ only the heteronuclear lines are visible. Thus by sweeping the photoassociation frequency over a range of detunings from the bare Rydberg line and measuring the relative strengths of different lines the relative atomic densities present in the mixture can be extracted. As the different lines can be associated with distinct radial distances, the relative strengths of different lines can also yield information about spatial correlations. The simple model presented here is generally applicable to all systems studied in this article since different perturbers will always, through their hyperfine structure and/or different scattering properties, have binding energies associated with a given Rydberg state that are distinguishable from homonuclear binding energies.

\section{Conclusion}
This paper has provided the low-energy phase shifts and sets of relevant empirical parameters necessary to compute accurate Rydberg molecule potential curves for all combinations of alkali atoms. The sample of potentials depicted here reveal the wide variety of novel molecular states present in heteronuclear Rydberg molecules, many of which are far more favorable to optical excitation than corresponding homonuclear states. Finally, since molecular states associated with different pertubers lead to markedly different spectroscopic signatures, the formation probability of heteronuclear molecules in an ultracold mixture depends strongly on the relative densities of atoms in the mixture, and is sensitive to these densities over a large range of distances much greater than typical atom-atom scattering lengths.   

The main aspects of this study can be readily extended to other atomic species, such as the alkaline-earth atoms or lanthanides, mirroring current work in quantum degenerate gases. In the case of the lanthanides, this requires \textit{ab initio} calculations of scattering phase shifts at low energy. Likewise, calculations \cite{BartschatSadeghpour} of these phase shifts in the alkaline earth atoms can be experimentally refined via spectroscopy. The phase shifts calculated here for the alkalis could be improved further with a fully relativistic treatment, and the model describing the density dependence of heteronuclear and homonuclear lines could be improved by including a more precise description of the line shapes, such as that given in Ref. \cite{OttGroup}.

\begin{acknowledgments}
I am grateful for many discussions with C. H. Greene, particularly regarding the phase shift calculations, and for his invaluable guidance regarding two-electron systems. I thank C. Fey and F. Hummel for their hospitality during a visit to Hamburg and for many enlightening discussions, and also P. Giannakeas and P\'{e}rez-R\'{i}os for their careful reading of the manuscript and many helpful suggestions. I acknowledge support from the NSF under Grant No. PHY1607180 and the MPI-PKS visitors program. 
\end{acknowledgments}
%\bibliography{D:/NewDropbox/Dropbox/Research/Prelim_documents/Thesis_bib}

\end{document}